%% file: 2024_Halldestam_DREAM_SPI_paper.tex
\documentclass{jpp}
\input{preamble.tex}

\shorttitle{Modelling SPI in AUG}
\title{Reduced kinetic modelling of shattered pellet injection in ASDEX Upgrade}

\shortauthor{P. Halldestam}
\author{
	P.~Halldestam\aff{1}\corresp{\email{peter.halldestam@ipp.mpg.de}},
  	P.~Heinrich\aff{1}, 
	G.~Papp\aff{1}, 
	M.~Hoppe\aff{2},
	M.~Hoelzl\aff{1}, 
	I.~Pusztai\aff{3}, 
	O.~Vallhagen\aff{3}, 
	R.~Fischer\aff{1}, 
	F.~Jenko\aff{1}, 
	the ASDEX Upgrade Team\aff{a}, 
	\and the EUROfusion Tokamak Exploitation Team\aff{b} 
}
\affiliation{
	\scriptsize
	\aff{1}Max Planck Institute for Plasma Physics, Garching, Germany 
	\aff{2}Department of Electrical Engineering, KTH Royal Institute of Technology, Stockholm, Sweden 
	\aff{3}Department of Physics, Chalmers University of Technology, G\"oteborg, Sweden
	\aff{a}See the author list of H.~Zohm \textit{et al.} (2024) Nucl.~Fusion
	\aff{b}See the author list of E.~Joffrin \textit{et al.} (2024) Nucl.~Fusion
}

\begin{document}

\maketitle
	
\begin{abstract}
Plasma-terminating disruptions represent a critical outstanding issue for reactor-relevant tokamaks.
ITER will use shattered pellet injection (SPI) as its disruption mitigation system to reduce heat loads, vessel forces, and to suppress the formation of runaway electrons. 
In this paper we demonstrate that reduced kinetic modelling of SPI is capable of capturing the major experimental trends in ASDEX Upgrade SPI experiments, such as dependence of the radiated energy fraction on neon content, or the current quench dynamics. 
Simulations are also consistent with the experimental observation of no runaway electron generation with neon and mixed deuterium-neon pellet composition.
We also show that statistical variations in the fragmentation process only have a notable impact on disruption dynamics at intermediate neon doping, as was observed in experiments.
\end{abstract}
	

\section{Introduction}
\label{sec:introduction}

\noindent
One of the main issues threatening the success of future reactor-scale tokamaks is disruptions.
It is the sudden loss of confinement, where the plasma rapidly ejects a large portion of its energy content onto the first wall, which exposes the device to excessive mechanical stresses, heat loads, and can lead to the formation of a runaway electron beam \citep{Hender2007, Breizman2019}.
Unmitigated disruptions could potentially cause severe damage to the device, and modelling such events is thus crucial to assess the effectiveness of various mitigation techniques and to optimise termination scenarios.

Shattered pellet injection (SPI) employs hydrogen isotopes (protium or deuterium) with minority neon admixture, frozen into cryogenic pellets, which are launched into the vessel at speeds of several hundred meters per second.
Upon entry, the singular pellet is broken into smaller pieces by a shattering unit to increase pellet material ablation and assimilation into the plasma. 
The exact pellet composition, penetration velocity, and the size distribution of pellet fragments has major impact on the material assimilation \citep{Jachmich_2023_EPS} and disruption mitigation efficiency \citep{Heinrich2025,Tang2024,Patel2024}. 
Modelling is crucial to understand the complex physics and to optimise the termination scenarios. 

The ITER baseline disruption mitigation system will be based on SPI due to its ability to promptly inject material into the plasma to thermally radiate its stored energy, reduce electromagnetic loads on the surrounding conducting structures, and to inhibit the generation of runaway electrons \citep{Hollmann2014, Lehnen2015, Jachmich2022, Vallhagen2024a}.  
As the mitigation of the various disruption consequences, however, poses conflicting criteria, finding an optimised scheme is not straightforward requiring experimental and theoretical input. One such conflict could be if different SPI penetration speeds and fragment distributions were required for mitigation thermal loads, vessel forces, or runaway electrons.
In support of its development, a highly flexible SPI system was installed at the ASDEX Upgrade (AUG) tokamak for the 2021/2022 SPI experimental campaign to investigate the impact of different fragment size and velocity distributions on the disruption dynamics \citep[see][]{Dibon2023, Heinrich2024}. The AUG system has so far been using deuterium as the primary hydrogen isotope. We do not expect major differences in the disruption dynamics between protium and deuterium SPI from the atomic physics point of view.

High-fidelity magnetohydrodynamic (MHD) simulations are routinely applied to disruption modelling, e.g.~with \JOREK~\citep{Hoelzl2021, Nardon2020, Hu2021, Tang2024} or \NIMROD~\citep{Sovinec2004, Kim2019}, representing the high physics fidelity---high numerical cost---end of the spectrum.
To carry out large parameter scans, aid design decisions, and help focus expensive simulations where most needed, reduced models are also employed. 
In this paper, the one-dimensional fluid model of the \DREAM~code (Disruption Runaway Electron Analysis Model, by \cite{Hoppe2021}) is used. 

This code has been used for simulations of ITER \citep{Pusztai2022, Lier2023, Ekmark2024, Vallhagen2024a} as well as TCV \citep{Wijkamp2024}, JET \citep{Jarvinen2022}, DIII-D \citep{Izzo2022, Marini2024}, STEP \citep{Berger2022, Fil2024}, and SPARC \citep{Izzo2022, Tinguely2023}. 
In this paper we demonstrate that \DREAM~can explain experimental trends of major disruption characteristics at AUG, such as the evolution of the plasma current $\Ip$ or the radiated energy fraction $\frad$.

The plume of fragments used as input for the modelling is generated by sampling fragment size and velocity distributions that depend on various injection parameters, such as the pellet neon fraction $\fNe$ and the injection speed $\vinj$, using the Parks model \citep{Parks2016} as implemented by \cite{Gebhart2020a}.
Since pellet fragmentation is a statistical process, we simulate an ensemble of several realisations for each parameter set, to assess the impact that statistical variation in the fragment distribution has on the resulting disruption dynamics (sensitivity analysis).

The physics model used is detailed in sections \ref{sec:physics-model} and \ref{sec:enhanced-transport}, followed by a description on the fragment sampling in section \ref{sec:sampling-fragments}.
An overview of the simulation setup and the AUG SPI discharges selected for the experimental comparison is presented in section \ref{sec:simulation-setup} and \ref{sec:selected-cases}, respectively.
A demonstrative scan in the pellet neon fraction is detailed in section \ref{sec:neonscan}, and the experimental comparison in section \ref{sec:experimental-comparison}. 
The results and limitations of the model are discussed and concluded in section \ref{sec:discussion-conclusion}.

\newpage
\section{Methods}

\subsection{Physics model}
\label{sec:physics-model}

\noindent
A one-dimensional fluid model of the plasma available in the \DREAM~code is applied.
This is a finite-volume numerical simulation framework that has been developed for tokamak disruption modelling, with a particular focus on the runaway electron dynamics.
It evolves the plasma current self-consistently, along with the thermal bulk of electrons, and charge states of the relevant ionic species -- all in a realistic toroidally symmetric magnetic geometry.
In this section, the relevant equations and physical assumptions are described in detail.

\subsubsection*{Plasma current}
\label{sec:plasma-current}

\noindent
The plasma current density $\jtot$ consists of an Ohmic component $\johm$, and a component $\jre$ carried by runaway electrons.
Other non-inductively driven currents, such as the bootstrap current, are not included in this model as these are expected to convert to Ohmic current during a disruption.
The total current density is related to the poloidal magnetic flux $\psip$, as described by Ampère's law
\begin{equation}
	\label{eq:amperes-law}
	2\pi\mu_0\fsa{\vb{B}\vdot\grad{\varphi}}\frac{\jtot}{B}
	=\frac{1}{V'}\pdv{r}V'\fsa{\frac{\abs{\grad{r}}^2}{R^2}}\pdv{\psip}{r},
\end{equation}
where $\fsa{\cdot}$ denotes flux surface average, $\vb{B}$ the magnetic field, $R$ the major radius, and $V'$ is the spatial Jacobian ($\dd{V}=V'\dd{r}$ being the volume between two infinitesimally close flux surfaces).
These depend on the particular magnetic equilibrium $\vb{B}(r, \theta)$ used in the simulation setup, which is treated as static in this model.
Note that radial coordinate $r$ is defined as the distance on the midplane from the magnetic axis $R=\Rm$ to the corresponding flux surface on the low-field side.

The poloidal magnetic flux $\psip$ is evolved using a modified version of Faraday's law of induction
\begin{equation}
	\label{eq:faradays-law}
	\pdv{\psip}{t}
	=-2\pi\frac{\fsa{\vb{E}\vdot\vb{B}}}{\fsa{\vb{B}\vdot\grad{\varphi}}}
	+\mu_0\pdv{\psit}\psit\hypres\pdv{\psit}\frac{\jtot}{B}.
\end{equation}
This is the loop voltage, of which the first term on the right-hand side is proportional to the electric field parallel to the magnetic field driving the Ohmic component of the current according to Ohm's law $\johm/B=\cond\fsa{\vb{E}\vdot\vb{B}}/\fsa{B^2}$.
The parallel electric conductivity is modelled as outlined in \cite{Redl2021}.
It is a neoclassical correction of the Spitzer conductivity $\cond\propto\Tcold^{3/2}$ accounting for trapped particle effects at arbitrary collisionality.

The second term in equation \eqref{eq:faradays-law}, derived by \cite{Boozer1986}, involves a second-order derivative in the toroidal magnetic flux $\psit=(2\pi)^{-1}\int V'\dd{r}\fsa{\vb{B}\cdot\grad{\varphi}}$, which is used as a radial coordinate.
Given the relation in equation \eqref{eq:amperes-law}, this term is in fact a fourth-order radial derivative, and thus describes the hyperdiffusion of $\psip$.
Whenever $\hypres$ is greater than zero, this term acts to relax $\jtot/B$ while conserving the helicity content of the plasma. 
This leads to a transient increase in $\Ip$ as the current density radially redistributes in what is referred to as the ``$\Ip$-spike''. 
This emulates the effects that the break-up of magnetic field lines has on the current density, and is used to model the enhanced transport event during the thermal quench phase of a disruption with strong field stochastisation.
This is explained in more detail in section \ref{sec:enhanced-transport}.

As boundary condition for the plasma edge flux $\psip(a)$, we couple it to an ideally conducting wall at $r=b>a$.
By doubly integrating equation \eqref{eq:amperes-law} over the vacuum region $a<r<b$, in which $\jtot=0$, we can relate the plasma edge flux to that of the wall $\psi(b)=\psi(a)+\Mab\Ip$, where $\Mab=\mu_0\Rm\ln(b/a)$ is the edge-wall mutual flux inductance taken in the cylindrical limit $a/R\to0$.
We consider a wall that is perfectly conducting, meaning no electric field is induced in the wall and thus the wall flux remains constant in time.

In addition to the Ohmic component, the plasma current can also contain a component driven by runaway electrons.
Without resolving the electron distribution in momentum space, the runaway electrons are assumed to travel at the speed of light $c$, and their contribution to the total plasma current density is given by $\jre=ec\nre$, with $e$ being the elementary charge.
The evolution of the runaway density $\nre$ also includes diffusive transport with $\Dre=1\,\rm m^2/s$, and a set of source terms accounting for the runaway electron generation mechanisms.
These include Dreicer generation, which is calculated by a neural network that has been trained on kinetic simulations \citep{Hesslow2019b, Ekmark2024}; hot-tail generation, as calculated in section 4.2 of \cite{Svenningsson2020}; and a model for the avalanche growth rate, which accounts for the effects of partially screened nuclei in non-ideal plasmas \citep{Hesslow2019a}.
A negligible generation of runaways is expected for the plasma scenarios considered in this work, but the respective terms are included for completeness as these models are readily available in \DREAM~and are relatively inexpensive computationally. 
Runaway electrons were so far not observed in ASDEX Upgrade SPI experiments with neon and deuterium-neon injections. 
Our simulations help explain that this is due to insufficient generation, rather than due to runaway seed losses preventing avalanche.

\subsubsection*{Ions and SPI dynamics}

\noindent
Ions and neutral atoms are characterised both by a temperature $\Ti$, and a set of densities $\nij{j}$ for every species $i$ with atomic number $Z_i$, and charge state $j=0,\dots Z_i$.
The densities are evolved according to
\begin{align}
	\label{eq:charge-states}
	\begin{split}
		\pdv{\nij{j}}{t}=\ionis{j-1}&\nij{j-1}\ncold-\ionis{j}\nij{j}\ncold
		+\recom{j+1}\nij{j+1}\ncold-\recom{j}\nij{j}\ncold\\
		&\quad+\delta_{0j}\sum_{k=1}^{\Nfrag}\Gk \frac{\delta(r-r_k)}{4\pi r^2\Rm}
		+\frac{1}{V'}\pdv{r}V'\left(\ionadv\nij{j}+\iondiff\pdv{\nij{j}}{r}\right).
	\end{split}
\end{align}
The first four terms describe ionisation and recombination processes, for which the coefficient $\ionis{j}$ denotes the rate of ionisation between the two charge states $j\to j+1$, and $\recom{j}$ the rate of recombination between $j\to j-1$.
These coefficients are extracted from the OpenADAS database \citep{OpenADAS}.

The fifth term describes a source into the neutral state $\nij{0}$ of the injected material, as the fragments travel along their trajectories $r_k(t),\, k=1,\dots\Nfrag$, depositing ablated material.
The rate at which a fragment decreases in diameter $d_k(t)$ is $\dot{d}_k=-2\Gk\Mmolar/(\pi d_k^2 \rho N_{\rm A})$, where $\Mmolar$ is the pellet molar mass, $\rho$ the pellet mass density, and $N_{\rm A}$ Avogadro's number.
The rate of the deposition of neutrals per unit time $\Gk$ is calculated using a version of the Neutral Gas Shielding (NGS) model, first derived by \cite{Parks1978} and further extended upon by \cite{Parks2017} to allow for mixed deuterium-neon pellets.
The NGS model provides the scaling law $\Gk\propto\lambda\ncold^{1/3}\Tcold^{5/3}d_k^{4/3}$, where $\lambda$ (along with $\Mmolar$ and $\rho$) is a function of the neon fraction of the pellet $\fNe=\NNe/(N_{\rm D_2}+\NNe)$.

The sixth and final term in equation \eqref{eq:charge-states} describes both advective and diffusive particle transport, for which the transport coefficients $\ionadv$ and $\iondiff$ are active during the thermal quench (similarly to $\hypres$).
More on this in section \ref{sec:enhanced-transport}.

For each ion species, a temperature $\Ti$, which is assumed to be same for all charge states, is evolved via collisional heat exchange with the bulk electron population and other ion species present in the plasma, according to
\begin{equation}
	\label{eq:ion-temperature}
	\frac{3}{2}\pdv{\ni\Ti}{t}
	=\sum_{i'}Q_{ii'}+Q_{ie},
\end{equation}
with the total density $\ni=\sum_{j=0}^{Z_i}\nij{j}$.
Here $Q_{\alpha\beta}$ is the collisional heat exchange between Maxwellian species
$\alpha$ and $\beta$, which is is proportional to their temperature difference, as further detailed in \cite{Hoppe2021}.

\vspace{.1cm}
\subsubsection*{Thermal electrons}
\noindent

The evolution of the electron temperature $\Tcold$ is modelled using an energy balance equation
\begin{align}
	\label{eq:energy-balance}
	\begin{split}
		\frac{3}{2}\pdv{\ncold\Tcold}{t}&=
		\frac{\johm}{B}\fsa{\vb{E}\vdot\vb{B}}
		-\ncold\sum_i\sum_{j=0}^{Z_i-1}\Lij\nij{j}
		+\Pion\\
		&\qquad+\frac{\jre}{B}\Ec\fsa{B}
		+\sum_i Q_{ei}
		+\frac{1}{V'}\pdv{r}V'\frac{3\ncold}{2}\heatdiff\pdv{\Tcold}{r},
	\end{split}
\end{align}
in which the first term on the right-hand side describes Ohmic heating, and the second radiative cooling due to line radiation, recombination radiation and bremsstrahlung. 
Integrating this term over the plasma volume yields the total radiated power, which we will denote as $\Prad$.
Typically, this energy balance is dominated by Ohmic heating and radiative cooling, both of which are nonlinear functions of the electron temperature.
The rates of energy loss $\Lij(\ncold, \Tcold)$ for neon are retrieved from OpenADAS \citep{OpenADAS}.
Deuterium is instead assumed to be opaque to Lyman radiation, which reduces the radiative losses at temperatures below $10\,\rm eV$; for this the loss rates are extracted from the AMJUEL database \citep{AMJUEL}.

The third term, $\Pion$, represents the change in potential energy due to the electron-ion binding energy removed by the electron bulk during ionisation and added during recombination.
With the ionisation threshold energies $\Delta W_i^{(j)}$ obtained from the NIST database \citep{NIST}, this contribution to the energy balance is calculated by adding up the terms $\Delta W_i^{(j)}(\recom{j}-\ionis{j})\nij{j}\ncold$ over all charge states.

Collisional heat transfer from the runaway electrons is accounted for with the fourth term, where $\Ec=\ncold e^3\ln{\Lambda}/(4\pi\varepsilon_0^2 m_e c^2)$.
The fifth term is the collisional heat transfer between electrons and ions $Q_{ei}=-Q_{ie}$, and the final term describes diffusive heat transport, with $\heatdiff$ being the heat diffusivity.
A baseline value of $\heatdiff=1\,\rm m^2/s$ is used throughout.

\subsection{Enhanced transport event}
\label{sec:enhanced-transport}

\noindent
To emulate the effect of the magnetic field stochastisation during the thermal quench that causes the thermal energy loss, an enhanced radial transport of heat, particles and poloidal magnetic flux is applied at a time $\trelax$.
As previously done in ITER studies by \cite{Vallhagen2024a}, we take $\trelax$ to be the time when the electron temperature drops below $10\,\rm eV$ somewhere inside of the resonant $q=2$ surface.
The choice of this particular criterion for triggering the enhanched transport event is motivated by the fact that the conductivity at $10\,\rm eV$ is sufficiently low to likely trigger a resistive MHD instability at the resonant $q=2$ flux surface on a millisecond timescale.

In the simulation the transport event is emulated by increasing any transport coefficient $\Y \in \{ \hypres, \ionadv, \iondiff, \heatdiff \}$ from their baseline value $\Y_0$ up to some maximum value $\Ymax$, which then exponentially decays over a timescale of $\tdecay=1\,\rm ms$, according to
\begin{equation}
	\label{eq:enhanced-transport}
	\Y(t)
	=\Y_0
	+(\Ymax-\Y_0)\exp\left(-\frac{t-\trelax}{\tdecay}\right)\Theta(t-\trelax),
\end{equation}
where $\Theta$ denotes the Heaviside function. 
In particular, this concerns the hyperdiffusivity $\hypres$ in equation \eqref{eq:faradays-law}, the ion and neutral particle advection and diffusion coefficients $\ionadv$ and $\iondiff$ in equation \eqref{eq:charge-states}, and the electron heat diffusivity $\heatdiff$ in the energy balance equation \eqref{eq:energy-balance}.
We omit the enhanced transport of runaway electrons, as all simulated cases considered here yield already negligible runaway currents.
We have conducted a sensitivity study for the impact of $\hypresmax$ on current quench evolution in the range of $\hypresmax=[1\times10^{-7}-1\times10^{-3}]\,\rm Wb^2 m/s$.
By selecting $\hypresmax=6.3\times10^{-5}\,\rm Wb^2 m/s$, we observe an increase in the plasma current during the $\Ip$-spike similar to the experiment ($\Delta\Ip\sim20\,\rm kA$ for the cases studied here). The impact of $\hypresmax$ in this range on the radiated energy fraction is between $\pm2\%$ and $\pm 5\%$, depending on the pellet composition.
The maximum value of the other transport coefficients are set to $\ionadvmax=-100\,\rm m/s$, $\iondiffmax=100\,\rm m^2/s$, and $\heatdiffmax=100\,\rm m^2/s$.
Similar values were used in ASTRA simulations of massive gas injection scenarios in AUG by \cite{Linder2020} to reproduce the evolution of line-integrated electron densities from experiment.

\subsection{Fragment sampling}
\label{sec:sampling-fragments}

\noindent 
The generation of the fragment plume is performed in two steps: a rejection sampling of fragment sizes, utilising an analytical distribution function $f(d)$ for the characteristic fragment diameter $d$ following the pellet break-up; and then assigning velocities $\vb{v}$ to the individual fragments.
This results in a set of tuples $\{(d_k, \vb{v}_k)\}$, with $k=1,\dots\Nfrag$.

The fragment size probability distribution stems from a statistics-based model for the shattering of brittle materials, initially derived in the context of exploding munitions by \cite{Mott1943}, and then more recently by \cite{Parks2016} in the context of shattered cryogenic pellets.
This has subsequently been correlated with relevant injection parameters to experimentally measured fragment distributions \citep{Gebhart2020a, Gebhart2020b}, which is used in this paper.
For a cylindrical pellet of diameter $D$ and length $L$ injected with an impact velocity $\vperp=\vinj\sin\shatterangle$ (component of the injection velocity $\vinj$ normal to the shatter plane at an angle $\shatterangle$), this fragmentation model yields the characteristic size distribution
\begin{equation} 
	\label{eq:fragment-distribution} 
	f(d)
	=\alpha d \bessel(\beta d). 
\end{equation} 
Here, $\alpha=\XR/D$, $\beta=\XR/LC$, and $\XR=\vperp^2/\vthres^2$, the latter of which is the ratio of the impact energy and the threshold energy required to initiate fracture in the pellet.
$\bessel$ is the modified Bessel function of the second kind accounting for the energy propagation through the pellet.  
The material parameters $C$ and $\vthres$ depend on the neon-deuterium content ratio $\fNe$ of the pellet, both of which are obtained from \cite{Gebhart2020b}.  
Fragments are sampled from \eqref{eq:fragment-distribution} in sequence until the sum of all fragment volumes exceeds the pellet volume, after which the final fragment is shrunk such that the total volume equals that of the pellet.  
In doing so, $\Nfrag$ will depend on the specific random seed used during the sampling process.

The next step concerns assigning velocities to each fragment.
For simplicity, the magnitude and direction are sampled independently from one another and uncorrelated to the corresponding fragment's size.  
The velocity magnitude $\vfrag$ is sampled from a normal distribution, where the mean fragment speed $\vfragmean$ is assumed to be given by the average of the injection speed $\vinj$ and its component parallel to the shattering surface at an angle of $\shatterangle=12.5^\circ$, i.e.~$\vfragmean=\vinj(1+\cos\shatterangle)/2$, and the spread is set to be $\vfragspread=20\,\%$ \citep[see][]{Peherstorfer2022}.  
Finally, each fragment is assigned a direction in which it traverses through the plasma at constant speed that is sampled uniformly within a circular cone with an opening angle of $20^\circ$ \citep{Peherstorfer2022}.

\subsection{Simulation setup}
\label{sec:simulation-setup}

\begin{figure}
	\centering 
	\vspace{.5cm}
	\includegraphics[width=.95\linewidth]{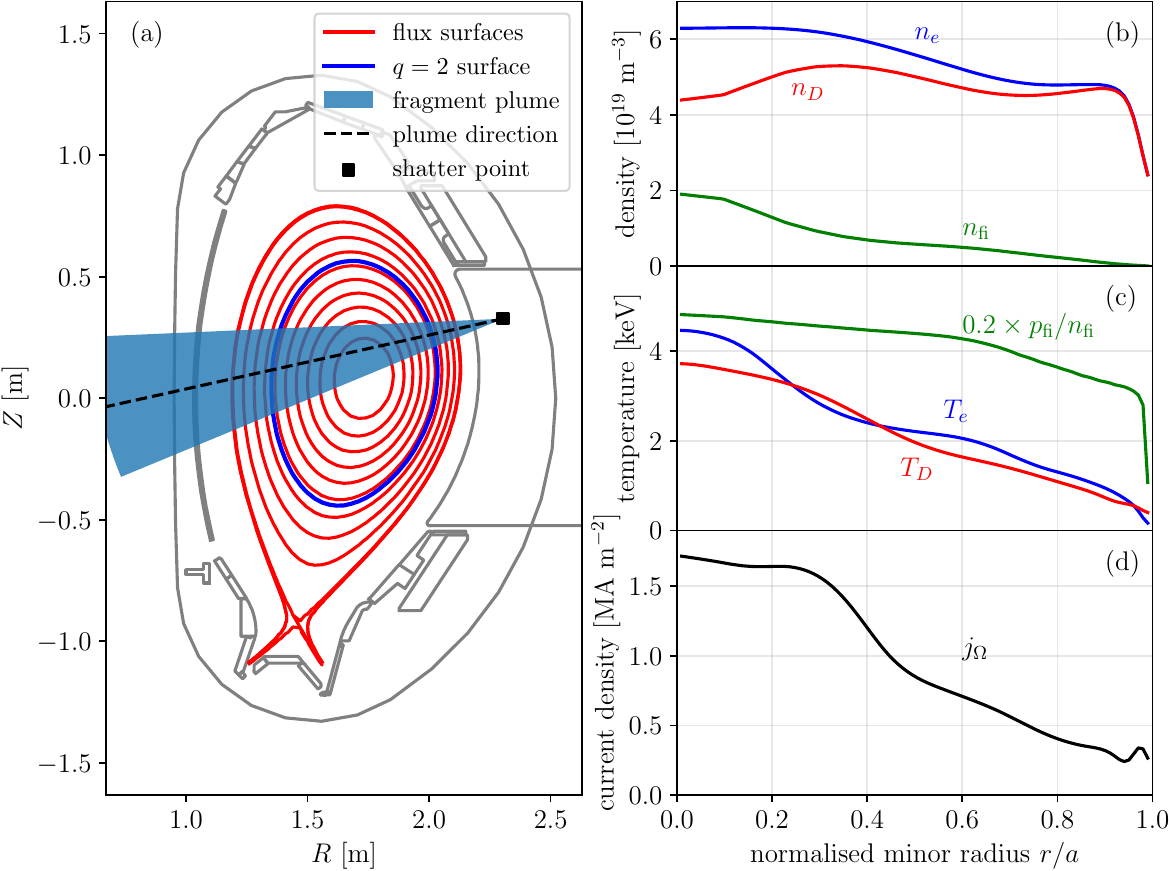}
	\caption{
Input data based on the reference AUG discharge \#40655, used in all simulations presented. 
An illustration (a) shows the flux surface geometry and the SPI plume, within which all fragments travel in straight lines originating from the shatter point.
The particular shatter head considered in this study has a square cross section with $22\,\rm mm$ side lengths, and a shatter angle $\shatterangle=12.5^\circ$.
Flux surfaces are indicated in red, and the $q=2$ surface is marked in blue.  
The initial plasma profiles used in this work consist of (b) density and (c) temperature profiles of the bulk electrons (blue), thermal ions (red) and fast ions (green), and (d) the initial current density. 
	}
	\label{fig:inputs} 
\end{figure}

\noindent 
Each simulation presented in this paper uses initial conditions from the representative AUG discharge \#40655, which is an H-mode deuterium plasma with a major radius of the magnetic axis $\Rm=1.74\,\rm m$, minor radius $a=0.39\,\rm m$ (midplane low-field side distance from $\Rm$ to the last closed flux surface), and on-axis toroidal magnetic field $B_0=1.81\,\rm T$.  
Reconstructed magnetic equilibrium data and relevant initial plasma profiles are taken at $t=2.3\,\rm s$, which is approximately the time of the SPI trigger, all of which are obtained from the Integrated Data Analysis (IDA) framework \citep{Fischer2010, Fischer2016, Fischer2020}.
IDA applies Bayesian probability theory on the combined analysis of multiple diagnostics and physical modelling to obtain the most cohesive profile reconstructions and uncertainty estimates.  

The reconstruction of the initial electron density profile $\ncold$ is based on both deuterium cyanide laser (DCN) interferometry and Thomson scattering (TS) spectroscopy, and the electron temperature $\Tcold$ is based on the latter. 
The initial thermal ion temperature $\Ti=\TD$ is based on charge exchange recombination spectroscopy (CXRS).
Furthermore, neutral beam injection (NBI) provides a non-negligible population of super-thermal fast ions, for which IDA makes estimates of the fast ion density $\nfi$ and pressure $\pfi$ profiles based on kinetic modelling using the RABBIT code \citep{Weiland2018}.
With the fast ions being made up of ionised deuterium, and assuming the pre-SPI plasma is ideal, the quasi-neutrality criterion implies that the initial thermal ion density can be expressed as $\nD=\ncold-\nfi$.  
The total plasma pressure $\ncold\Tcold+\nD\TD+\pfi$ is then used to solve for the magnetic equilibrium in the Grad-Shafranov equation, which is done iteratively since these profiles themselves depend on the equilibrium \citep[see][]{Fischer2016}. 

In this model we do not resolve the momentum space distribution of any of the particle species involved.
Instead, the thermal and fast ions are incorporated into a single Maxwellian distribution, with density $\nD=\ncold$ and a rescaled temperature as to maintain the total plasma pressure.  
By doing this, it is effectively assumed that these fast ions instantaneously thermalise at the start of the simulation. 
It is also assumed that NBI is turned off at this point in time, meaning no further fast ions are introduced.

The corresponding flux surface geometry, with the SPI plume indicated, and the relevant plasma profiles are shown in figure \ref{fig:inputs}.  
Note that the total current density from experiment includes bootstrap current and the NBI driven current, which is here incorporated into the Ohmic current, as shown in figure \ref{fig:inputs}d.  
This is motivated by the fact that the bootstrap current quickly becomes Ohmic as a result of the relaxation of the pressure profile during the enhanced transport event.  
The NBI driven current is turned off by the control system as soon as insufficient absorption is detected, and since we do not model the complex control system logic, this contribution to the total current is omitted.

The radial coordinate is discretised on a grid with $n_r=11$ points, and for the time coordinate we use an adaptive timestep based on the assumption that the shortest timescales are those of ionisation $\tau_{\rm ionis}\sim\abs{\partial\log\ncold/\partial t}^{-1}$.
It sets the next timestep according to $\Delta t=\min(\Delta t_{\rm max}, \alpha \tau_{\rm ionis})$, where we set $\Delta t_{\rm max}=10^{-6}\, \rm s$ as the maximum allowed timestep and $\alpha=10^{-3}$.
With this setup, depending on the pellet neon content, the typical wall-clock time for the simulations considered here is between 5-15 minutes on a computer running on an AMD EPYC 9354 processor.

\begin{table} 
	\centering 
	\caption{ 
SPI parameters used in the simulations of the considered AUG discharges, being the pellet neon fraction $\fNe$, number of injected neon atoms $\NNe$, pellet length $L$ and diameter $D$, and pellet injection speed	$\vinj$. 
All injections use the $\shatterangle=12.5^\circ$ shatter head with a square cross-section -- corresponding to guide tube 3 (GT3) in the 2022 experimental setup -- as indicated in figure \ref{fig:inputs}a.
The final column shows the mean and standard deviation in the number of fragments generated using these parameters, based on 40 samplings of the fragment size distribution \eqref{eq:fragment-distribution}.
	}
	\vspace{.5cm} 
	\begin{tabular}{llllllll}
		\# 	& $\fNe$ [\%] 	& $\NNe$ & $D$ ($L$) [mm]	& $\vinj$ [m/s]	& $\Nfrag$	\\ 
		\midrule
		40679	& 100	& $1.93\times10^{22}$	& 8.0 (9.0) & 259.4	& $49.5\pm15.6$	\\
		40677	& 40	& $6.49\times10^{21}$	& 8.0 (9.5)	& 374.8	& $67.4\pm23.3$ \\ 
		40949	& 10	& $1.67\times10^{21}$	& 7.9 (11.0)& 214.6	& $5.9\pm2.6$	\\ 
		40732	& 1.25	& $1.79\times10^{20}$	& 7.9 (9.7)	& 232.1	& $5.9\pm2.7$	\\
		40737	& 0.17	& $2.24\times10^{19}$	& 8.0 (8.7)	& 231.9 & $6.3\pm2.9$	\\ 
		40738	& 0		& $0$					& 8.1 (10.7)& 219.0	& $5.5\pm2.9$	\\
	\end{tabular}
	\vspace{.5cm} 
	\label{tab:discharges} 
\end{table}

\subsection{Experimental cases selected for comparison}
\label{sec:selected-cases}

\noindent 
The SPI system installed for the 2022 AUG SPI experimental campaign is highly flexible, with three different shatter heads, of which two have a square cross section at a $\shatterangle=12.5^\circ$ and $25^\circ$ shatter angle, respectively, and the third has a circular cross section at $\shatterangle=25^\circ$.  
With two different shatter angles it is possible to isolate the normal impact velocity $\vperp=\vinj\sin\shatterangle$, which sets the fragment size distribution as given in equation \eqref{eq:fragment-distribution}, and the velocity distribution of the fragments $\vfrag$.  
A detailed description of the SPI system is provided by \cite{Dibon2023, Heinrich2024}.

In the 2022 AUG SPI campaign, about 180 SPI shots were carried out with this same H-mode scenario, of which \#40655 was selected as the representative discharge. To represent a scan in injected neon quantity, we have chosen a small subset of discharges from the same scenario, each of which have relevant plasma parameters (such as plasma stored energy or pedestal top temperature) within $\lesssim 5\%$ of the reference shot.
In this paper, all the considered cases are pellets injected via the square shatter head with $\shatterangle=12.5^\circ$ (GT3).  
With input conditions all based on the reference AUG discharge \#40655, as described in section \ref{sec:simulation-setup}, only the injection parameters are modified when modeling the various discharges, mainly the fraction of neon $\fNe$ in the injected pellet. 
An overview of the different setups can is shown in table~\ref{tab:discharges}.  
The values for the pellet length $L$, diameter $D$ and injection speed $\vinj$ are based on image processing analysis by \cite{Peherstorfer2022}. 
Note, that the 40\% neon case has a higher injection velocity than the rest. There were no experiments executed in 2022 with 40\% neon and a matching velocity, but we nevertheless wished to include a neon fraction case between 10\% and 100\% in the experimental comparison.
For each discharge we run 40 otherwise identical simulations with varying random seeds, resulting in a variation in the number of fragments sampled between the simulations.  
The last column in table \ref{tab:discharges} includes the average and standard deviation in the number of fragments.

\section{Simulation results and experimental comparison}

\begin{figure}
	\centering 
	\vspace{.5cm}
	\includegraphics[width=.95\linewidth]{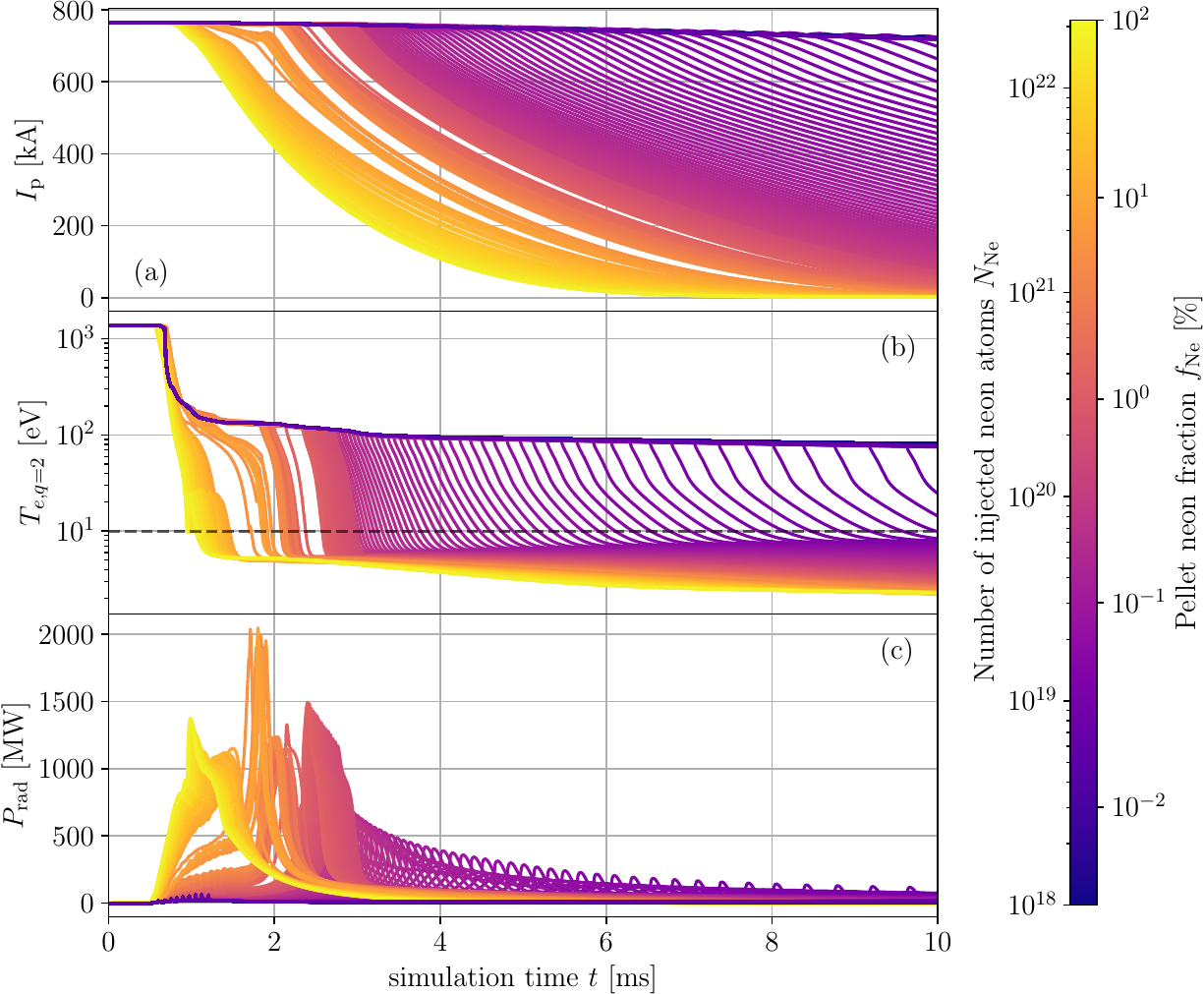}
	\caption{ 
Scan in the amount of injected neon, showcasing the evolution of the (a) plasma current, (b) electron temperature at the resonant $q=2$ surface, and the (c) total radiated power.  
The dashed line in (b) indicates the temperature threshold of $10\,\rm eV$ at which we trigger the enhanced transport event.
The scan consists of 300 individual simulations, each with fragments sampled from equation \eqref{eq:fragment-distribution} which depends on $\fNe$.
	} 
	\label{fig:neonscan}
\end{figure}

\subsection{Pellet neon fraction scan} 
\label{sec:neonscan}

\noindent 
To gauge the effect the pellet neon fraction $\fNe$ (or equivalently, the number of injected neon atoms, $\NNe$) has in this dynamic system, we scan in the range between $100\,\%$ down to $0.0001\,\%$ in search of trends.
Figure \ref{fig:neonscan} shows the plasma current, electron temperature at the $q=2$ surface, and the total radiated power, as functions of time for different values of $\fNe$.
The initial temperature drop in all cases is due to dilution cooling as the density is rapidly increased.
With more neon introduced to the system, more energy is lost via radiation, leading to a more rapid drop in temperature, which in turn causes an earlier onset time for the enhanced transport event triggered.
Since the conductivity scales as $\cond\propto\Tcold^{3/2}$, a stronger temperature drop with an increasing amount of injected neon accelerates the current drop during the current quench.

For some cases with less than $5\,\%$ trace amounts of neon, or $\NNe\approx1.5\times10^{21}$, rather than moving in from the edge towards the $q=2$ surface, the thermal collapse below $10\,\rm eV$ is initiated in the core region. 
When the fragments pass $R=\Rm$ and reach the high-field side, they again deposit material on flux surfaces they already passed on the low-field side.
For cases with neon fractions $\fNe\lesssim0.001\,\%$, or $\NNe\lesssim3\times10^{17}$, a thermal collapse does not occur, and the plasma therefore does not disrupt. 

The reason for the deep fragment penetration in the simulations is twofold. Less neon in the simulations will allow the fragments to penetrate deeper because it takes more time before the thermal collapse occurs due to less radiative cooling. At the same time, with smaller neon fractions the Parks model predicts fewer fragments that take longer to ablate, further improving their penetration.
In this manuscript we do not isolate the effects of pellet composition from the number of fragments. A similar scan in the pellet composition is done by \cite{Patel2024}, in which the number of fragments are kept fixed at 200 across all neon fractions. This same phenomenon of an inside-out collapse at lower neon fractions is also observed, suggesting it is in fact the lower radiative cooling that allows for a deeper fragment penetration, and not only the low number of fragments generated. Core penetration of SPI is also observed experimentally in the ultra high speed video recordings of AUG SPI experiments~\citep{Patel2023}.

In the context of runaway electron generation, the maximum simulated runaway current is approximately $5\,\rm A$, occuring in the cases with majority neon admixture.
This is mainly attributed to the larger runaway seed generated by the hot-tail mechanism due to the more pronounced thermal quench.
However, the induced electric field following this drop in electron temperature is insufficient to produce a significant amount of runaway electrons from this seed population via the avalanche mechanism.

\begin{figure}
	\centering
	\vspace{.5cm}
	\includegraphics[width=.85\linewidth]{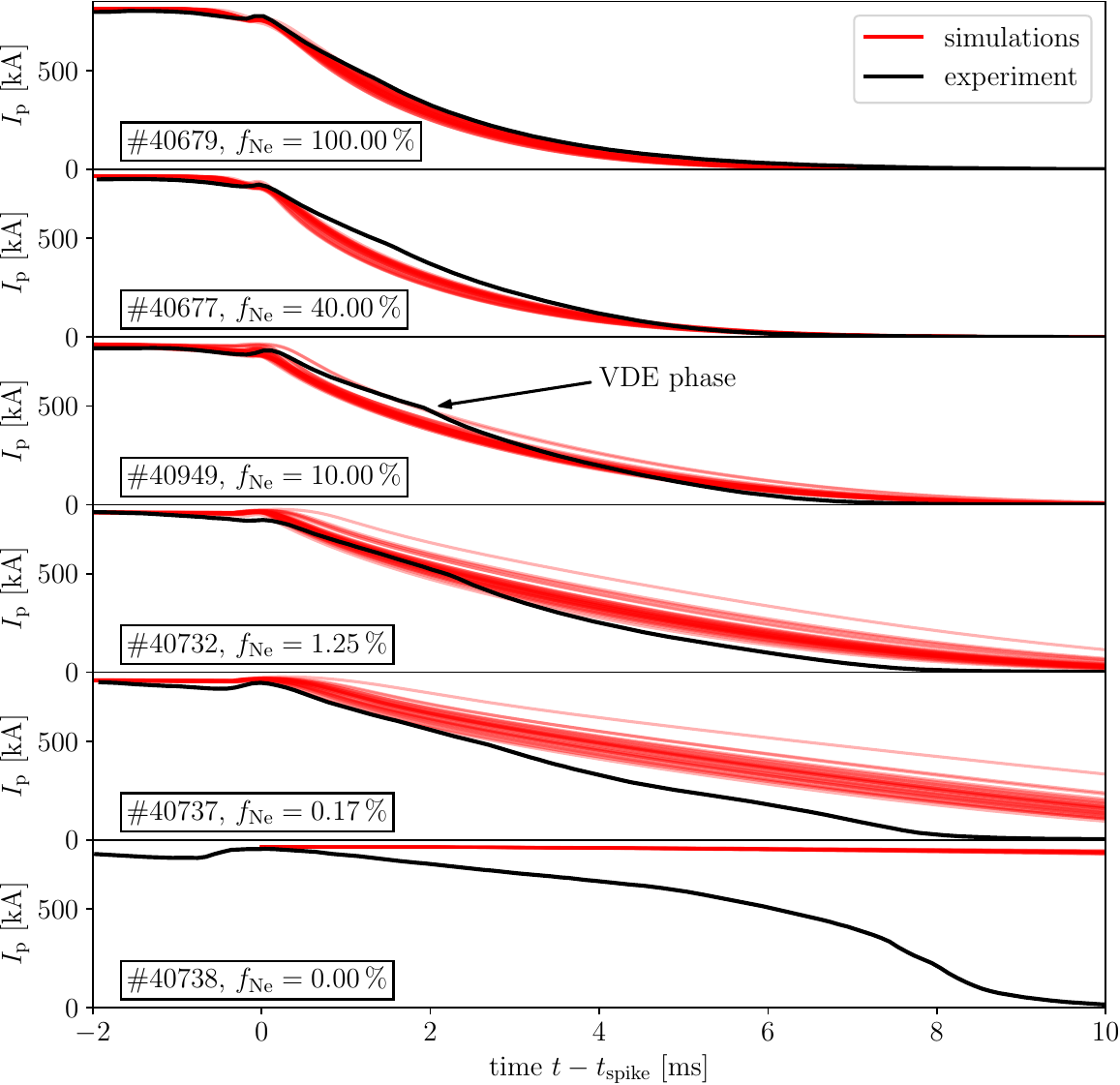} 
	\caption{
Comparison of the plasma current evolution for varying amounts of injected neon in experiment measured data (black) and simulations (red), for which each case is run 40 times using different random seeds while sampling fragments.
Except for the pure deuterium injection, a good agreement with experimental plasma currents is observed during the initial current quench evolution (100\,\% to 80\,\% of $\Ip$) prior to the onset of the VDE phase.
	}
	\label{fig:currents} 
\end{figure}

\subsection{Plasma current evolution} 
\label{sec:experimental-comparison}

\noindent
Simulated plasma currents of six selected injection scenarios are compared with experimental measurements in figure \ref{fig:currents}.
Each case was run 40 times using different realisations of the fragment sampling procedure.
For the  cases with a finite amount of injected neon, the simulations generally agree well with experiment.
Note that since the magnetic field is kept static in the simulations, no runaway scrap-off is accounted for \citep{Vallhagen2024b}, and effects of vertical displacement events (VDEs) are not captured.
These VDEs may explain the kink seen in the experimental data at about halfway down the current quench. 
In addition, we do not model any control system action, as the AUG discharge control system is too complex to implement in such a simulation, and a feed-forward implementation of the actions executed would lead to a departure from self-consistent modeling of the plasma evolution.

The last column in table \ref{tab:discharges} displays the average number and standard deviation of fragments sampled for the six injection scenarios.
The fragment size distribution \eqref{eq:fragment-distribution} has a dependence on the neon fraction and the impact velocity $\vperp$ (or $\vinj$, as we fix the shatter angle $\shatterangle=12.5^\circ$) that determines the size of the fragments, or equivalently the total number of fragments sampled.

In discharge \#40738 no neon is introduced into the plasma, and in the simulations the radiated losses are not sufficient to initialise the enhanced transport event, meaning that the simulated plasma does not disrupt. This was expected in the modelling due to the relatively low radiation of deuterium.
In the experiment we observed a relatively long quench of the plasma current. 
Full-sized pure deuterium pellets injected into the AUG SPI scenario plasma typically lead to a disruption, either via an inadvertent control system action~\citep{Sieglin2020,Heinrich2025}, or as is suspected in this case, due to background impurity radiation (mainly tungsten from the first wall). 
The presence of background impurities provides additional loss channels through line radiation, enabling the decrease in temperature below $10\,\rm eV$ within the $q=2$ flux surface, and thus triggering the enhanced transport event.
However, it was also observed that 100\,\% D$_2$ injections containing fewer than $N_{\rm D_2}=10^{22}$ deuterium molecules do not disrupt the H-mode scenario used for the SPI experiments \citep{Heinrich2025}.

\subsection{Radiated energy fraction}

\begin{figure}
	\centering
	\vspace{.5cm}
	\includegraphics[width=.65\linewidth]{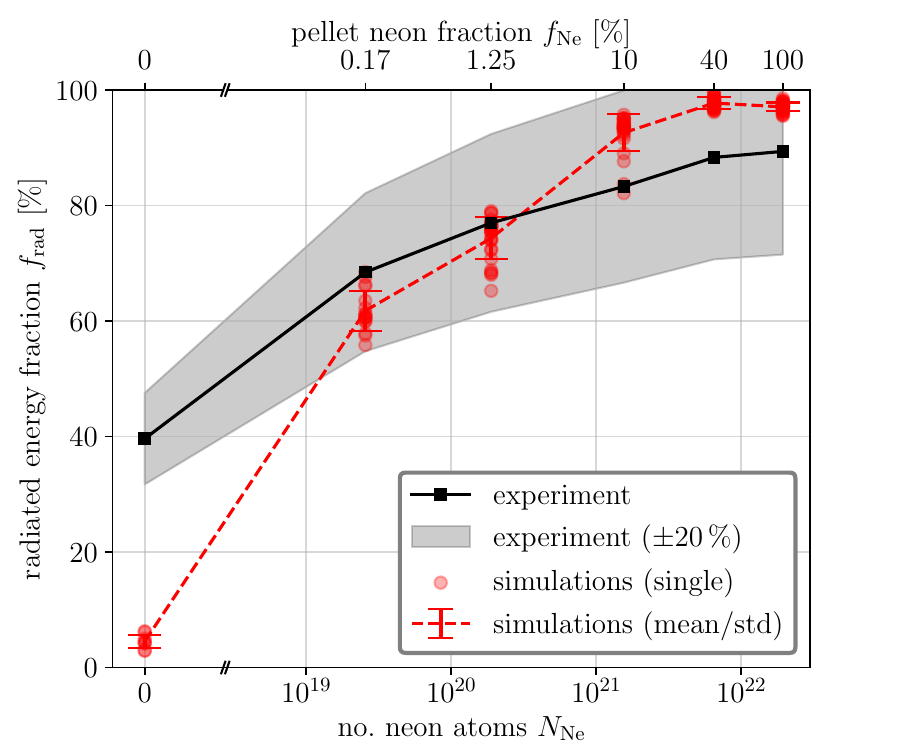} 
	\caption{
Radiated energy fraction $\frad$ from simulations (red) and experimental estimates (black) as functions of the amount of injected neon.  
The scatter in the simulation data comes from generating fragment samples using 40 different random seeds in otherwise identical simulations.
For the experimental estimates, a relative uncertainty of $20\,\%$ (gray) is assumed \citep{Heinrich2025}.
	} 
	\label{fig:frad}
\end{figure}

\noindent
The radiated energy fraction $\frad$ shows how much of the total available plasma energy drop is lost during a disruption via radiation, as opposed to conduction.  
It is calculated using the formula by \cite{Lehnen2011} and \cite{Sheikh2020} 
\begin{equation}
	\label{eq:frad}
	\frad=\frac{\Wrad}{\Wth+\Wmag+\Wheat-\Wcoup}, 
\end{equation} 
where $\Wrad$ is the radiated energy, $\Wth$ and $\Wmag$ are respectively the initial thermal and poloidal magnetic energy, $\Wheat$ is external heating, and $\Wcoup$ is the energy coupled into the surrounding conducting structure.
For the simulations, the radiated energy is calculated as $\Wrad=\int_{\tstart}^{\tend}\dd{t}\Prad$, where the time $\tstart$ is when the first fragment reaches the plasma, and $\tend$ is taken to be at the end of the radiation peak, when the derivative of $\Prad$ approaches zero, as was done for the experimental estimates \citep{Heinrich2025}.
The initial thermal energy $\Wth$ is the combined energy of all thermal particles, i.e.~$\Wth=3/2(\ncold\Tcold+\sum_i\ni\Ti)$, evaluated at $t=\tstart$.  
Typically in AUG experiments, the heating systems automatically turn off due to insufficient absorption shortly after the fragments enter the plasma, and will therefore contribute to the total energy of the system for a brief time. In the experimental evaluation of $\frad$ we correct for this by subtracting $\Wheat$, which accounts for the external heating of the plasma between the arrival of fragments and shut-off of the external heating systems \citep{Heinrich2025}. In the current simulation setup, we can simply assume that all heating system are effectively turned off at $t=\tstart$, and thus we set $\Wheat=0$. 
Furthermore, the energy coupled to the surrounding structure $\Wcoup=0$, since we assume a perfectly conducting wall as described in section \ref{sec:plasma-current}. 

Figure \ref{fig:frad} shows a comparison between the radiated energy fraction from simulations and experimental estimates, as described by \cite{Heinrich2025}.
The calculation involves interpolating foil bolometry measurements of the radiated power in different sectors of the tokamak, and integrating it over an interval of time similarly determined as in the simulation calculations.
The experimental values have an estimated relative uncertainty of about $20\,\%$.

Another key difference between simulation and experiment for 100\,\% D$_2$ injection, apart from the plasma current, is $\frad$.
Otherwise, the simulated radiated energy fraction aligns well with experiment results, both in magnitude and its scaling with the neon fraction.
While within the experimental uncertainty margins, $\frad$ for the highest neon fractions is close to 100\,\% as opposed to the 90\,\% observed in the experiments.
There is yet no definite conclusion for the reason of the saturation of the experimentally observed $\frad$ around 90\,\%. 
A potential cause for this may be the incorrect estimation of conducted and coupled energy losses. However, both of these are difficult to self-consistently model in a 1D simulation. 

Compared to 3D MHD simulations \citep{Tang2024} we see that the 1D \DREAM~simulations somewhat overestimate $\frad$. While \DREAM~provides a better match than \JOREK~for the experiments at low neon doping above about 1\,\% neon content \JOREK~simulations show a radiated energy fraction closer to the experimental values than \DREAM.

\section{Discussion and conclusion}
\label{sec:discussion-conclusion}

\noindent
We applied a one-dimensional fluid model of SPI-induced disruptions in the ASDEX Upgrade tokamak, dynamically evolving the plasma current, thermal bulk of electrons, and ion charge state densities.
This numerically inexpensive model was demonstrated to yield good results when comparing to experimental data, capturing the general trend in the plasma current evolution and radiated energy fraction when varying the amount of injected neon. 
We also observe no significant runaway generation, as in the experiments. 

A notable exception from the close agreement is the case of 100\,\% deuterium injection, where the experiments do disrupt eventually (if $N_{\rm D_2}\gtrsim10^{22}$), typically due to the combined effect of background impurity radiation, reaching the density limit, vertical instability, or control system action.
These aspects are not yet modelled here.
In particular, the discrepancy between the simulated radiated energy fraction and the experimental estimates is primarily attributed to the absence of background impurities in the simulations. 
These impurities would enhance radiative energy losses, reducing the temperature sufficiently to trigger the observed enhanced transport event.
The inclusion of tungsten as a background impurity in the plasma as well as the material deposition shift due to plasmoid drifts is proposed for future investigations.

A probabilistic method for generating individual fragment sizes and velocities is presented, the former of which is based on fragmentation model first proposed by \cite{Parks2016} and modified by \cite{Gebhart2020a, Gebhart2020b}.  With the present setup based on Parks, statistical variations in this sampling procedure provide an uncertainty quantification and are shown to only have a notable impact on the disruption dynamics for trace amounts of injected neon.
The same is also observed experimentally \citep{Heinrich2025}.
This means, that while thermal load mitigation would be difficult to optimise by selecting appropriate size distributions, it is also not a factor to worry about when determining the optimal SPI parameters for e.g. material assimilation for runaway electron suppression.

A potential cause for discrepancies between simulation and experiment is that sampling from the Parks fragment size distribution has been shown by \cite{Peherstorfer2022} to differ from the experimentally measured fragment size distributions at AUG.
The Peherstorfer analysis was, however, limited to $4\,\rm mm$ diameter pellets with typically low neon content, most of which do not cause disruptions at AUG, and thus are not suitable for experimental validation of disruption simulations. Furthermore, we wanted to first compare results obtained with the Parks model to the experiments, as this is widely used in the literature and thus a validation exercise was considered useful for the community. 
Recent peridynamic modelling by \cite{Lee2024} has shown a promising match between modelled and measured fragment size distributions. Furthermore, reconstructed fragment size and velocity distributions from fast camera videos using a deep learning-based algorithm~\citep{Illerhaus2024} are becoming available. Investigating the effect of these different fragment distributions on the disruption dynamics is planned for a future study.
Comparisons to 3D MHD simulations can help to inform model parameters and joint studies are planned in the future with fast 1D scans complemented by more numerically expensive non-linear 3D MHD studies of the most relevant scenarios \citep{Tang2024}.

\section*{Acknowledgements}
\noindent 
The authors are grateful to S.~Jachmich, J.~Artola, W.~Tang and A.~H.~Patel for the fruitful discussions.
This work has been carried out within the framework of the EUROfusion Consortium, funded by the European Union via the Euratom Research and Training Programme (Grant Agreement No 101052200 — EUROfusion). 
Views and opinions expressed are however those of the author(s) only and do not necessarily reflect those of the European Union, the European Commission, or the ITER Organisation.
Neither the European Union nor the European Commission can be held responsible for them.
The ASDEX-Upgrade SPI project has been implemented as part of the ITER DMS Task Force programme. 
The SPI system and related diagnostics have received funding from the ITER Organisation under contracts IO/20/CT/43-2084, IO/20/CT/43-2115, IO/20/CT/43-2116.

\section*{Declaration of interest}
\noindent
The authors report no conflict of interest.

\newpage	
\bibliographystyle{jpp}
\bibliography{refs}

\end{document}

%% file: preamble.tex
\usepackage{physics}
\usepackage{graphicx}
\usepackage{color}
\usepackage{booktabs}
\usepackage{float}

\usepackage[utf8]{inputenc}
\usepackage[T1]{fontenc}
\usepackage{amsmath}
\usepackage{url}

\renewcommand{\vb}{\boldsymbol}

\newcommand{\DREAM}{\textsc{Dream}}
\newcommand{\JOREK}{\textsc{Jorek}}
\newcommand{\NIMROD}{\textsc{Nimrod}}

\newcommand{\XR}{X_{\rm R}}
\newcommand{\vinj}{v_{\rm inj}}
\newcommand{\vperp}{v_\perp}
\newcommand{\vthres}{v_{\rm thres}}
\newcommand{\bessel}{K_0}

\newcommand{\vfrag}{v_{\rm frag}}
\newcommand{\vfragmean}{\overline{v}_{\rm frag}}
\newcommand{\vfragspread}{\Delta\vfrag/\vfragmean}
\newcommand{\Nfrag}{N_{\rm frag}}
\newcommand{\shatterangle}{\theta_{\rm s}}

\newcommand{\NNe}{N_{\rm Ne}}
\newcommand{\fNe}{f_{\rm Ne}}

\newcommand{\frad}{f_{\rm rad}}
\newcommand{\Wrad}{W_{\rm rad}}
\newcommand{\Wcoup}{W_{\rm c}}
\newcommand{\Wth}{W_{\rm th}}
\newcommand{\Wmag}{W_{\rm mag}}
\newcommand{\Wheat}{W_{\rm ext}}
\newcommand{\tstart}{t_{\rm start}}
\newcommand{\tend}{t_{\rm end}}

\newcommand{\Rm}{R_0}

\newcommand{\fsa}[1]{\left\langle#1\right\rangle}

\newcommand{\ncold}{n_e}
\newcommand{\Tcold}{T_e}
\newcommand{\johm}{j_\Omega}
\newcommand{\cond}{\sigma}

\newcommand{\Pion}{\mathcal{P}_{\rm ion}}

\newcommand{\Prad}{P_{\rm rad}}
\newcommand{\Lij}{L_i^{(j)}}

\newcommand{\Ti}{T_i}
\renewcommand{\ni}{n_i}
\newcommand{\nij}[1]{n_i^{(#1)}}

\newcommand{\ionis}[1]{\mathcal{I}_i^{(#1)}}
\newcommand{\recom}[1]{\mathcal{R}_i^{(#1)}}
\newcommand{\nD}{n_{\rm D}}
\newcommand{\TD}{T_{\rm D}}
\newcommand{\nfi}{n_{\rm fi}}
\newcommand{\pfi}{p_{\rm fi}}

\newcommand{\jtot}{j_\parallel}
\newcommand{\psip}{\psi_{\rm p}}
\newcommand{\psit}{\psi_{\rm t}}

\newcommand{\Ip}{I_{\rm p}}
\newcommand{\Mab}{M_{\rm ab}}

\newcommand{\jre}{j_{\rm re}}
\newcommand{\nre}{n_{\rm re}}
\newcommand{\Dre}{D_{\rm re}}
\newcommand{\Ec}{E_{\rm c}}

\newcommand{\Y}{\mathcal{Y}}
\newcommand{\Ymax}{\mathcal{Y}_{\rm max}}
\newcommand{\trelax}{t_{\rm relax}}
\newcommand{\tdecay}{\tau_{\rm decay}}
\newcommand{\hypres}{\Lambda}
\newcommand{\iondiff}{D_i}
\newcommand{\ionadv}{A_i}
\newcommand{\heatdiff}{\chi_e}
\newcommand{\hypresmax}{\hypres_{\rm max}}
\newcommand{\iondiffmax}{D_{i,\rm max}}
\newcommand{\ionadvmax}{A_{i,\rm max}}
\newcommand{\heatdiffmax}{\chi_{e,\rm max}}

\newcommand{\Gk}{\mathcal{G}_k}
\newcommand{\Mmolar}{\mathcal{M}}

%% file: refs.bib
@inproceedings{Jachmich_2023_EPS,
	author = {Jachmich, S and Lehnen, M and Papp, G and Kruezi, U and Bernert, M and Bock, A and Cerovsky, J and Dibon, M and Ficker, O and Fiorucci, D and Heinrich, P and Kocsis, G and Kurzan, B and de Marne, P and Peherstorfer, T and Sheikh, U and Eich, T and Hobirk, J and the ASDEX Upgrade Team},
	booktitle = {Europhysics Conference Abstracts},
	volume = {47A},
	pages = {O2.103},
	title = {Shattered Pellet Injection experiments at {ASDEX-Upgrade} for design optimisation of the {ITER} Disruption Mitigation System},
	address = {Bordeaux, France},
	url = {https://lac913.epfl.ch/epsppd3/2023/html/Orals/O2_103_Jachmich.pdf},
	isbn = {111-22-33333-44-5},
	year = {2023}}

@online{AMJUEL,
	author={D. Reiter},
	title={The data file {AMJUEL}: Additional Atomic and Molecular Data for {EIRENE}},
	year={2020},
	url={https://www.eirene.de/old_eirene/html/amjuel.html}
}

@article{Berger2022, 
	title = {Runaway dynamics in reactor-scale spherical tokamak disruptions}, 
	volume = {88}, 
	DOI = {10.1017/S0022377822001209}, 
	number = {6}, 
	journal = {Journal of Plasma Physics}, 
	author = {E. Berger and I. Pusztai and S. L. Newton and M. Hoppe and O. Vallhagen and A. Fil and T. Fülöp}, 
	year = {2022}, 
	pages = {905880611}
}

@article{Boozer1986,
	title={Ohm’s law for mean magnetic fields}, 
	volume={35}, 
	DOI={10.1017/S0022377800011181},
	journal={Journal of Plasma Physics}, 
	author={Boozer, A. H.}, 
	year={1986}, 
	pages={133–139},
	url = {https://doi.org/10.1017/S0022377800011181}
}

@article{Breizman2019,
	doi = {10.1088/1741-4326/ab1822},
	url = {https://dx.doi.org/10.1088/1741-4326/ab1822},
	year = {2019},
	month = {jun},
	publisher = {IOP Publishing},
	volume = {59},
	number = {8},
	pages = {083001},
	author = {Boris N. Breizman and Pavel Aleynikov and Eric M. Hollmann and Michael Lehnen},
	title = {Physics of runaway electrons in tokamaks},
	journal = {Nuclear Fusion},
}

@article{Kim2019,
    author = {Kim, Charlson C. and Liu, Yueqiang and Parks, Paul B. and Lao, Lang L. and Lehnen, Michael and Loarte, Alberto},
    title = {Shattered pellet injection simulations with {NIMROD}},
    journal = {Physics of Plasmas},
    volume = {26},
    number = {4},
    pages = {042510},
    year = {2019},
    month = {04},
    issn = {1070-664X},
    doi = {10.1063/1.5088814},
    url = {https://doi.org/10.1063/1.5088814},
    eprint = {https://pubs.aip.org/aip/pop/article-pdf/doi/10.1063/1.5088814/14020971/042510\_1\_online.pdf},
}

@article{Dibon2023,
    author = {Dibon, M. and de Marne, P. and Papp, G. and Vinyar, I. and Lukin, A. and Jachmich, S. and Kruezi, U. and Muir, A. and Rohde, V. and Lehnen, M. and Heinrich, P. and Peherstorfer, T. and Podymskii, D. and ASDEX Upgrade Team},
    title = {Design of the shattered pellet injection system for {ASDEX Upgrade}},
    journal = {Review of Scientific Instruments},
    volume = {94},
    number = {4},
    pages = {043504},
    year = {2023},
    month = {04},
    issn = {0034-6748},
    doi = {10.1063/5.0141799},
    url = {https://doi.org/10.1063/5.0141799},
}

@article{Ekmark2024, 
	title={Fluid and kinetic studies of tokamak disruptions using Bayesian optimization}, 
	volume={90}, 
	DOI={10.1017/S0022377824000606}, 
	number={3}, 
	journal={Journal of Plasma Physics}, 
	author={Ekmark, I. and Hoppe, M. and Fülöp, T. and Jansson, P. and Antonsson, L. and Vallhagen, O. and Pusztai, I.}, 
	year={2024}, 
	pages={905900306}
}

@article{Fil2024,
	doi = {10.1088/1741-4326/ad73e9},
	url = {https://dx.doi.org/10.1088/1741-4326/ad73e9},
	year = {2024},
	month = {sep},
	publisher = {IOP Publishing},
	volume = {64},
	number = {10},
	pages = {106049},
	author = {A. Fil and L. Henden and S. Newton and M. Hoppe and O. Vallhagen},
	title = {Disruption runaway electron generation and mitigation in the Spherical Tokamak for Energy Production ({STEP})},
	journal = {Nuclear Fusion},
}

@article{Fischer2010,
	author = {R. Fischer and C. J. Fuchs and B. Kurzan and W. Suttrop and E. Wolfrum},
	title = {Integrated Data Analysis of Profile Diagnostics at {ASDEX Upgrade}},
	journal = {Fusion Science and Technology},
	volume = {58},
	number = {2},
	pages = {675--684},
	year = {2010},
	publisher = {Taylor \& Francis},
	doi = {10.13182/FST10-110},
	url = {https://doi.org/10.13182/FST10-110},
}

@article{Fischer2016,
	author = {R. Fischer and A. Bock and M. Dunne and J. C. Fuchs and L. Giannone and K. Lackner and P. J. McCarthy and E. Poli and R. Preuss and M. Rampp and M. Schubert and J. Stober and W. Suttrop and G. Tardini and M. Weiland},
	title = {Coupling of the Flux Diffusion Equation with the Equilibrium Reconstruction at {ASDEX Upgrade}},
	journal = {Fusion Science and Technology},
	volume = {69},
	number = {2},
	pages = {526--536},
	year = {2016},
	publisher = {American Nuclear Society},
	doi = {10.13182/FST15-185},
	URL = {https://doi.org/10.13182/FST15-185}
}

@article{Fischer2020,
	author = {R. Fischer and L. Giannone and J. Illerhaus and P. J. McCarthy and R. M. McDermott and ASDEX Upgrade Team},
	title = {Estimation and Uncertainties of Profiles and Equilibria for Fusion Modeling Codes},
	journal = {Fusion Science and Technology},
	volume = {76},
	number = {8},
	pages = {879--893},
	year = {2020},
	publisher = {American Nuclear Society},
	doi = {10.1080/15361055.2020.1820794},
	URL = {https://doi.org/10.1080/15361055.2020.1820794},
}

@article{Gebhart2020a,
  	author={Gebhart, T. E. and Baylor, L. R. and Meitner, S. J.},
  	journal={IEEE Transactions on Plasma Science}, 
  	title={Experimental Pellet Shatter Thresholds and Analysis of Shatter Tube Ejecta for Disruption Mitigation Cryogenic Pellets}, 
  	year={2020},
  	volume={48},
  	number={6},
  	pages={1598-1605},
  	doi={10.1109/TPS.2019.2957968},
	url = {https://doi.org/10.1109/TPS.2019.2957968}
}

@article{Gebhart2020b,
	author = {T. E. Gebhart and L. R. Baylor and S. J. Meitner},
	title = {Shatter Thresholds and Fragment Size Distributions of Deuterium–Neon Mixture Cryogenic Pellets for Tokamak Thermal Mitigation},
	journal = {Fusion Science and Technology},
	volume = {76},
	number = {7},
	pages = {831--835},
	year = {2020},
	publisher = {American Nuclear Society},
	doi = {10.1080/15361055.2020.1812991},
	url ={https://doi.org/10.1080/15361055.2020.1812991},
}

@article{Heinrich2024,
	title = {Recipes for pellet generation and launching in the {ASDEX Upgrade SPI}},
	journal = {Fusion Engineering and Design},
	volume = {206},
	pages = {114576},
	year = {2024},
	issn = {0920-3796},
	doi = {https://doi.org/10.1016/j.fusengdes.2024.114576},
	url = {https://www.sciencedirect.com/science/article/pii/S0920379624004289},
	author = {P. Heinrich and G. Papp and P. {de Marné} and M. Dibon and S. Jachmich and M. Lehnen and T. Peherstorfer and I. Vinyar},
}

@article{Heinrich2025,
	doi = {10.1088/1741-4326/adcbc0},
	url = {https://dx.doi.org/10.1088/1741-4326/adcbc0},
	year = {2025},
	month = {apr},
	publisher = {IOP Publishing},
	volume = {65},
	number = {5},
	pages = {056036},
	author = {Heinrich, P. and Papp, G. and Jachmich, S. and Artola, J. and Bernert, M. and de Marné, P. and Dibon, M. and Dux, R. and Eberl, T. and Hobirk, J. and Lehnen, M. and Peherstorfer, T. and Schwarz, N. and Sheikh, U. and Sieglin, B. and Svoboda, J. and the ASDEX Upgrade Team and the EUROfusion Tokamak Exploitation Team},
	title = {Radiated energy fraction of SPI-induced disruptions at ASDEX Upgrade},
	journal = {Nuclear Fusion},
}

@article{Hender2007,
	doi = {10.1088/0029-5515/47/6/S03},
	url = {https://dx.doi.org/10.1088/0029-5515/47/6/S03},
	year = {2007},
	month = {jun},
	publisher = {},
	volume = {47},
	number = {6},
	pages = {S128},
	author = {T.C. Hender and J.C Wesley and J. Bialek and A. Bondeson and A.H. Boozer and R.J. Buttery and A. Garofalo and T.P Goodman and R.S. Granetz and Y. Gribov and O. Gruber and M. Gryaznevich and G. Giruzzi and S. Günter and N. Hayashi and P. Helander and C.C. Hegna and D.F. Howell and D.A. Humphreys and G.T.A. Huysmans and A.W. Hyatt and A. Isayama and S.C. Jardin and Y. Kawano and A. Kellman and C. Kessel and H.R. Koslowski and R.J. La Haye and E. Lazzaro and Y.Q. Liu and V. Lukash and J. Manickam and S. Medvedev and V. Mertens and S.V. Mirnov and Y. Nakamura and G. Navratil and M. Okabayashi and T. Ozeki and R. Paccagnella and G. Pautasso and F. Porcelli and V.D. Pustovitov and V. Riccardo and M. Sato and O. Sauter and M.J. Schaffer and M. Shimada and P. Sonato and E.J. Strait and M. Sugihara and M. Takechi and A.D. Turnbull and E. Westerhof and D.G. Whyte and R. Yoshino and H. Zohm and the ITPA MHD, Disruption and Magnetic Control Topical Group},
	title = {Chapter 3: {MHD} stability, operational limits and disruptions},
	journal = {Nuclear Fusion},
}

@article{Hesslow2019a,
	doi={10.1088/1741-4326/ab26c2},
	url={https://dx.doi.org/10.1088/1741-4326/ab26c2},
	year={2019},
	month={jun},
	publisher={IOP Publishing},
	volume={59},
	number={8},
	pages={084004},
	author={L. Hesslow and O. Embréus and O. Vallhagen and T. Fülöp},
	title={Influence of massive material injection on avalanche runaway generation during tokamak disruptions},
	journal={Nuclear Fusion}
}

@article{Hesslow2019b,
	title={Evaluation of the {Dreicer} runaway generation rate in the presence of high-$Z$ impurities using a neural network}, 
	volume={85}, 
	DOI={10.1017/S0022377819000874}, 
	number={6}, 
	journal={Journal of Plasma Physics}, 
	author={Hesslow, L. and Unnerfelt, L. and Vallhagen, O. and Embreus, O. and Hoppe, M. and Papp, G. and Fülöp, T.}, 
	year={2019}, 
	pages={475850601}
}

@article{Hoelzl2021,
	doi = {10.1088/1741-4326/abf99f},
	url = {https://dx.doi.org/10.1088/1741-4326/abf99f},
	year = {2021},
	month = {may},
	publisher = {IOP Publishing},
	volume = {61},
	number = {6},
	pages = {065001},
	author = {M. Hoelzl and G.T.A. Huijsmans and S.J.P. Pamela and M. Bécoulet and E. Nardon and F.J. Artola and B. Nkonga and C.V. Atanasiu and V. Bandaru and A. Bhole and D. Bonfiglio and A. Cathey and O. Czarny and A. Dvornova and T. Fehér and A. Fil and E. Franck and S. Futatani and M. Gruca and H. Guillard and J.W. Haverkort and I. Holod and D. Hu and S.K. Kim and S.Q. Korving and L. Kos and I. Krebs and L. Kripner and G. Latu and F. Liu and P. Merkel and D. Meshcheriakov and V. Mitterauer and S. Mochalskyy and J.A. Morales and R. Nies and N. Nikulsin and F. Orain and J. Pratt and R. Ramasamy and P. Ramet and C. Reux and K. Särkimäki and N. Schwarz and P. Singh Verma and S.F. Smith and C. Sommariva and E. Strumberger and D.C. van Vugt and M. Verbeek and E. Westerhof and F. Wieschollek and J. Zielinski},
	title = {The {JOREK} non-linear extended {MHD} code and applications to large-scale instabilities and their control in magnetically confined fusion plasmas},
	journal = {Nuclear Fusion},
}

@article{Hollmann2014,
    author = {Hollmann, E. M. and Aleynikov, P. B. and Fülöp, T. and Humphreys, D. A. and Izzo, V. A. and Lehnen, M. and Lukash, V. E. and Papp, G. and Pautasso, G. and Saint-Laurent, F. and Snipes, J. A.},
    title = {Status of research toward the {ITER} disruption mitigation system},
    journal = {Physics of Plasmas},
    volume = {22},
    number = {2},
    pages = {021802},
    year = {2014},
    month = {11},
    issn = {1070-664X},
    doi = {10.1063/1.4901251},
    url = {https://doi.org/10.1063/1.4901251},
    eprint = {https://pubs.aip.org/aip/pop/article-pdf/doi/10.1063/1.4901251/16137172/021802\_1\_online.pdf},
}

@article{Hoppe2021,
	title = {{DREAM}: A fluid-kinetic framework for tokamak disruption runaway electron simulations},
	journal = {Computer Physics Communications},
	volume = {268},
	pages = {108098},
	year = {2021},
	issn = {0010-4655},
	doi = {https://doi.org/10.1016/j.cpc.2021.108098},
	url = {https://www.sciencedirect.com/science/article/pii/S0010465521002101},
	author = {Mathias Hoppe and Ola Embreus and Tünde Fülöp},
}

@article{Hu2021,
	doi = {10.1088/1741-4326/abcbcb},
	url = {https://dx.doi.org/10.1088/1741-4326/abcbcb},
	year = {2021},
	month = {jan},
	publisher = {IOP Publishing},
	volume = {61},
	number = {2},
	pages = {026015},
	author = {D. Hu and E. Nardon and M. Hoelzl and F. Wieschollek and M. Lehnen and G.T.A. Huijsmans and D. C. van Vugt and S.-H. Kim and JET contributors and JOREK team},
	title = {Radiation asymmetry and {MHD} destabilization during the thermal quench after impurity shattered pellet injection},
	journal = {Nuclear Fusion},
}

@article{Illerhaus2024,
	title = {Status of the Deep Learning-Based Shattered Pellet Injection Shard Tracking at {ASDEX Upgrade}},
	journal = {Journal of Fusion Energy},
	volume = {43},
	number = {14},
	year = {2024},
	doi = {10.1007/s10894-024-00406-x},
	url = {https://doi.org/10.1007/s10894-024-00406-x},
	author = {Illerhaus, J. and Treutterer, W. and Heinrich, P. and Miah, M. and Papp, G. and Peherstorfer, T. and Sieglin, B. and Toussaint, U.v. and Zohm, H. and Jenko, F. and the ASDEX Upgrade Team}
}

@article{Izzo2022,
	doi = {10.1088/1741-4326/ac83d8},
	url = {https://dx.doi.org/10.1088/1741-4326/ac83d8},
	year = {2022},
	month = {aug},
	publisher = {IOP Publishing},
	volume = {62},
	number = {9},
	pages = {096029},
	author = {V.A. Izzo and I. Pusztai and K. Särkimäki and A. Sundström and D.T. Garnier and D. Weisberg and R.A. Tinguely and C. Paz-Soldan and R.S. Granetz and R. Sweeney},
	title = {Runaway electron deconfinement in {SPARC} and {DIII-D} by a passive {3D} coil},
	journal = {Nuclear Fusion},
}

@article{Jachmich2022,
	doi = {10.1088/1741-4326/ac3c86},
	url = {https://dx.doi.org/10.1088/1741-4326/ac3c86},
	year = {2021},
	month = {dec},
	publisher = {IOP Publishing},
	volume = {62},
	number = {2},
	pages = {026012},
	author = {S. Jachmich and U. Kruezi and M. Lehnen and M. Baruzzo and L.R. Baylor and D. Carnevale and D. Craven and N.W. Eidietis and O. Ficker and T.E. Gebhart and S. Gerasimov and J.L. Herfindal and E. Hollmann and A. Huber and P. Lomas and J. Lovell and A. Manzanares and M. Maslov and J. Mlynar and G. Pautasso and C. Paz-Soldan and A. Peacock and L. Piron and V. Plyusnin and M. Reinke and C. Reux and F. Rimini and U. Sheikh and D. Shiraki and S. Silburn and R. Sweeney and J. Wilson and P. Carvalho and the JET Contributors},
	title = {Shattered pellet injection experiments at {JET} in support of the {ITER} disruption mitigation system design},
	journal = {Nuclear Fusion},
}

@article{Jarvinen2022,
	title = {Bayesian approach for validation of runaway electron simulations}, 
	volume = {88}, 
	DOI = {10.1017/S0022377822001210}, 
	number = {6}, 
	journal = {Journal of Plasma Physics}, 
	author = {Järvinen, A.E. and Fülöp, T. and Hirvijoki, E. and Hoppe, M. and Kit, A. and Åström, J.},
	year={2022}, 
	pages={905880612}
}

@Article{Lee2024,
  author    = {Lee, S.-J. and Madenci, E. and Na, Yong-Su and de Marné, P. and Dibon, M. and Heinrich, P. and Jachmich, S. and Papp, G. and Peherstorfer, T.},
  journal   = {Nuclear Fusion},
  title     = {Peridynamic modelling of cryogenic deuterium pellet fragmentation for shattered pellet injection in tokamaks},
  year      = {2024},
  issn      = {1741-4326},
  month     = aug,
  number    = {10},
  pages     = {106023},
  volume    = {64},
  doi       = {10.1088/1741-4326/ad69a3},
  publisher = {IOP Publishing},
  url       = {https://doi.org/10.1088/1741-4326/ad69a3}
}

@article{Lehnen2011,
	doi = {10.1088/0029-5515/51/12/123010},
	url = {https://dx.doi.org/10.1088/0029-5515/51/12/123010},
	year = {2011},
	month = {nov},
	publisher = {},
	volume = {51},
	number = {12},
	pages = {123010},
	author = {M. Lehnen and A. Alonso and G. Arnoux and N. Baumgarten and S.A. Bozhenkov and S. Brezinsek and M. Brix and T. Eich and S.N. Gerasimov and A. Huber and S. Jachmich and U. Kruezi and P.D. Morgan and V.V. Plyusnin and C. Reux and V. Riccardo and G. Sergienko and M.F. Stamp and JET EFDA contributors},
	title = {Disruption mitigation by massive gas injection in {JET}},
	journal = {Nuclear Fusion},
}

@article{Lehnen2015,
	title = {Disruptions in {ITER} and strategies for their control and mitigation},
	journal = {Journal of Nuclear Materials},
	volume = {463},
	pages = {39-48},
	year = {2015},
	issn = {0022-3115},
	doi = {10.1016/j.jnucmat.2014.10.075},
	url = {https://doi.org/10.1016/j.jnucmat.2014.10.075},
	author = {M. Lehnen and K. Aleynikova and P.B. Aleynikov and D.J. Campbell and P. Drewelow and N.W. Eidietis and Yu. Gasparyan and R.S. Granetz and Y. Gribov and N. Hartmann and E.M. Hollmann and V.A. Izzo and S. Jachmich and S.-H. Kim and M. Kočan and H.R. Koslowski and D. Kovalenko and U. Kruezi and A. Loarte and S. Maruyama and G.F. Matthews and P.B. Parks and G. Pautasso and R.A. Pitts and C. Reux and V. Riccardo and R. Roccella and J.A. Snipes and A.J. Thornton and P.C. {de Vries}},
}

@article{Lier2023,
	doi = {10.1088/1741-4326/acc4de},
	url = {https://dx.doi.org/10.1088/1741-4326/acc4de},
	year = {2023},
	month = {apr},
	publisher = {IOP Publishing},
	volume = {63},
	number = {5},
	pages = {056018},
	author = {A. Lier and G. Papp and Ph. W. Lauber and I. Pusztai and K. Särkimäki and O. Embreus},
	title = {The impact of fusion-born alpha particles on runaway electron dynamics in {ITER} disruptions},
	journal = {Nuclear Fusion},
}

@article{Linder2020,
	doi = {10.1088/1741-4326/ab9dcf},
	url = {https://dx.doi.org/10.1088/1741-4326/ab9dcf},
	year = {2020},
	month = {aug},
	publisher = {IOP Publishing},
	volume = {60},
	number = {9},
	pages = {096031},
	author = {O. Linder and E. Fable and F. Jenko and G. Papp and G. Pautasso and the ASDEX Upgrade team and the EUROfusion MST1 team},
	title = {Self-consistent modeling of runaway electron generation in massive gas injection scenarios in {ASDEX Upgrade}},
	journal = {Nuclear Fusion},
}

@article{Marini2024,
	doi = {10.1088/1741-4326/ad4db6},
	url = {https://dx.doi.org/10.1088/1741-4326/ad4db6},
	year = {2024},
	month = {jun},
	publisher = {IOP Publishing},
	volume = {64},
	number = {7},
	pages = {076039},
	author = {C. Marini and E.M. Hollmann and S.W. Tang and J.L. Herfindal and D. Shiraki and R.S. Wilcox and D. del-Castillo-Negrete and M. Yang and N. Eidietis and M. Hoppe},
	title = {Runaway electron plateau current profile reconstruction from synchrotron imaging and {Ar-II} line polarization angle measurements in {DIII-D}},
	journal = {Nuclear Fusion},
}

@article{Mott1943,
	title = {A Theory of Fragmentation},
	journal = {UK Ministry of Supply},
	author = {N. Mott and E. Linfoot},
	year = {1943}
}

@article{Nardon2020,
	doi = {10.1088/1741-4326/abb749},
	url = {https://dx.doi.org/10.1088/1741-4326/abb749},
	year = {2020},
	month = {oct},
	publisher = {IOP Publishing},
	volume = {60},
	number = {12},
	pages = {126040},
	author = {E. Nardon and D. Hu and M. Hoelzl and D. Bonfiglio and the JOREK team},
	title = {Fast plasma dilution in {ITER} with pure deuterium shattered pellet injection},
	journal = {Nuclear Fusion},
}

@online{NIST,
	author={A. Kramida and Y. Ralchenko and J. Reader and NIST ASD Team},
	title={{NIST} Atomic Spectra Database (ver. 5.8)},
	year={2020},
	url={https://www.nist.gov/pml/asb}
}

@online{OpenADAS,
	author={H.P. Summers},
	title={The {ADAS} user manual, version 2.6},
	year={2004},
	url={http://www.adas.ac.uk}
}

@article{Parks1978,
    author = {Parks, P. B. and Turnbull, R. J.},
    title = {Effect of transonic flow in the ablation cloud on the lifetime of a solid hydrogen pellet in a plasma},
    journal = {The Physics of Fluids},
    volume = {21},
    number = {10},
    pages = {1735-1741},
    year = {1978},
    month = {10},
    issn = {0031-9171},
    doi = {10.1063/1.862088},
    url = {https://doi.org/10.1063/1.862088},
    eprint = {https://pubs.aip.org/aip/pfl/article-pdf/21/10/1735/12480997/1735\_1\_online.pdf},
}

@article{Parks2016,
  	title={Modeling dynamic fracture of cryogenic pellets},
  	author={Parks, Paul},
  	year={2016},
  	institution={General Atomics, San Diego, CA (United States)},
 	journal={General Atomics technical report},
  	volume={GA},
  	number={A28352},
	url={https://doi.org/10.2172/1344852}
}

@online{Parks2017,
	title = {A Theoretical Model for the Penetration of a Shattered-Pellet Debris plume},
	author = {Parks, Paul},
	year = {2017},
	url = {https://tsdw.pppl.gov/Talks/2017/Lexar/Paul\%20PPPL\%20talk.pdf},
}

@article{Peherstorfer2022,
	title={Fragmentation Analysis of Cryogenic Pellets for Disruption Mitigation},
  	author={T. Peherstorfer},
  	journal={arXiv preprint},
	pages={arXiv:2209.01024},
  	year={2022},
	url = {https://arxiv.org/abs/2209.01024}
}

@article{Pusztai2022, 
	title={Runaway dynamics in tokamak disruptions with current relaxation}, 
	volume={88}, 
	DOI={10.1017/S0022377822000733}, 
	number={4}, 
	journal={Journal of Plasma Physics}, 
	author={Pusztai, István and Hoppe, Mathias and Vallhagen, Oskar}, 
	year={2022}, 
	pages={905880409}
}

@article{Redl2021,
    author = {Redl, A. and Angioni, C. and Belli, E. and Sauter, O. and ASDEX Upgrade Team and EUROfusion MST1 Team},
    title = {A new set of analytical formulae for the computation of the bootstrap current and the neoclassical conductivity in tokamaks},
    journal = {Physics of Plasmas},
    volume = {28},
    number = {2},
    pages = {022502},
    year = {2021},
    month = {02},
    issn = {1070-664X},
    doi = {10.1063/5.0012664},
    url = {https://doi.org/10.1063/5.0012664},
}

@article{Sheikh2020,
	doi = {10.1088/1741-4326/abb425},
	url = {https://dx.doi.org/10.1088/1741-4326/abb425},
	year = {2020},
	month = {oct},
	publisher = {IOP Publishing},
	volume = {60},
	number = {12},
	pages = {126029},
	author = {U.A. Sheikh and P. David and O. Ficker and M. Bernert and D. Brida and M. Dibon and B. Duval and M. Faitsch and M. Maraschek and G. Papp and G. Pautasso and C. Sozzi and the AUG team and EUROfusion MST1 team},
	title = {Disruption mitigation efficiency and scaling with thermal energy fraction on {ASDEX Upgrade}},
	journal = {Nuclear Fusion},
}

@article{Sieglin2020,
	title = {Rapid prototyping of advanced control schemes in {ASDEX Upgrade}},
	journal = {Fusion Engineering and Design},
	volume = {161},
	pages = {111958},
	year = {2020},
	issn = {0920-3796},
	doi = {https://doi.org/10.1016/j.fusengdes.2020.111958},
	url = {https://www.sciencedirect.com/science/article/pii/S0920379620305068},
	author = {B. Sieglin and M. Maraschek and O. Kudlacek and A. Gude and W. Treutterer and M. Kölbl and A. Lenz},
}

@article{Sovinec2004,
	title = {Nonlinear magnetohydrodynamics simulation using high-order finite elements},
	journal = {Journal of Computational Physics},
	volume = {195},
	number = {1},
	pages = {355-386},
	year = {2004},
	issn = {0021-9991},
	doi = {https://doi.org/10.1016/j.jcp.2003.10.004},
	url = {https://www.sciencedirect.com/science/article/pii/S0021999103005369},
	author = {C.R. Sovinec and A.H. Glasser and T.A. Gianakon and D.C. Barnes and R.A. Nebel and S.E. Kruger and D.D. Schnack and S.J. Plimpton and A. Tarditi and M.S. Chu},

}

@mastersthesis{Patel2023,
	title={Modelling of shattered pellet injection experiments on the
{ASDEX Upgrade} tokamak},
	author={A. H. Patel},
	school={Eindhoven University of Technology},
	year={2023},
	url = {https://arxiv.org/abs/2312.03462},
	note = {\url{https://arxiv.org/abs/2312.03462}}

}

@mastersthesis{Svenningsson2020,
	title={Hot-tail runaway electron generation in cooling fusion plasmas},
	author={I. Svenningsson},
	school={Chalmers university of technology},
	year={2020}
}

@article{Tang2024,
	title = {Non-linear shattered pellet injection modelling in {ASDEX Upgrade}},
	author = {W. Tang and M. Hoelzl and M. Lehnen and D. Hu and F. J. Artola and P. Halldestam and P. Heinrich and S. Jachmich and E. Nardon and G. Papp and A. Patel and the ASDEX Upgrade Team and the EUROfusion Tokamak Exploitation Team and the JOREK Team},
	journal = {Nuclear Fusion},
	number = {submitted},
	year = {2025},
	url = {https://arxiv.org/abs/2412.03112},
	note = "\url{https://arxiv.org/abs/2412.03112}"
}

@article{Patel2024,
	title = {Modelling of shattered pellet injection experiments on the {ASDEX Upgrade} tokamak},
	author = {A. Patel and A. Matsuyama and G. Papp and M. Lehnen and J. Artola and S. Jachmich and E. Fable and A. Bock and B. Kurzan and M. Hoelzl and W. Tang and M. Dunne and R. Fischer and P. Heinrich and the ASDEX Upgrade Team and the EUROfusion Tokamak Exploitation Team},
	journal = {Nuclear Fusion},
	number = {submitted},
	year = {2025},
	url = {https://arxiv.org/abs/2502.09357},
	note = "\url{https://arxiv.org/abs/2502.09357}"
}

@article{Tinguely2023,
	doi = {10.1088/1361-6587/acb083},
	url = {https://dx.doi.org/10.1088/1361-6587/acb083},
	year = {2023},
	month = {jan},
	publisher = {IOP Publishing},
	volume = {65},
	number = {3},
	pages = {034002},
	author = {R A Tinguely and I Pusztai and V A Izzo and K Särkimäki and T Fülöp and D T Garnier and R S Granetz and M Hoppe and C Paz-Soldan and A Sundström and R Sweeney},
	title = {On the minimum transport required to passively suppress runaway electrons in {SPARC} disruptions},
	journal = {Plasma Physics and Controlled Fusion},
}

@article{Vallhagen2024a,
	doi = {10.1088/1741-4326/ad54d7},
	url = {https://dx.doi.org/10.1088/1741-4326/ad54d7},
	year = {2024},
	month = {jun},
	publisher = {IOP Publishing},
	volume = {64},
	number = {8},
	pages = {086033},
	author = {O. Vallhagen and L. Hanebring and F.J. Artola and M. Lehnen and E. Nardon and T. Fülöp and M. Hoppe and S.L. Newton and I. Pusztai},
	title = {Runaway electron dynamics in {ITER} disruptions with shattered pellet injections},
	journal = {Nuclear Fusion},
}

@article{Vallhagen2024b,
	author = {O. Vallhagen and L. Hanebring and T. Fülöp and M. Hoppe and I. Pusztai},
	title= {Reduced modeling of scrape-off losses of runaway electrons during tokamak disruptions}, 
	journal = {Submitted to J. Plasma Phys.},
	year = {2024},
	doi = {arxiv.org/abs/2410.03512}
}

@article{Weiland2018,
	doi = {10.1088/1741-4326/aabf0f},
	url = {https://dx.doi.org/10.1088/1741-4326/aabf0f},
	year = {2018},
	month = {jul},
	publisher = {IOP Publishing},
	volume = {58},
	number = {8},
	pages = {082032},
	author = {M. Weiland and R. Bilato and R. Dux and B. Geiger and A. Lebschy and F. Felici and R. Fischer and D. Rittich and M. van Zeeland and the ASDEX Upgrade Team and the Eurofusion MST1 Team},
	title = {{RABBIT}: Real-time simulation of the NBI fast-ion distribution},
	journal = {Nuclear Fusion},
}

@article{Wijkamp2024,
	doi = {10.1088/1741-4326/ad0dd7},
	url = {https://dx.doi.org/10.1088/1741-4326/ad0dd7},
	year = {2023},
	month = {nov},
	publisher = {IOP Publishing},
	volume = {64},
	number = {1},
	pages = {016021},
	author = {T.A. Wijkamp and M. Hoppe and J. Decker and B.P. Duval and A. Perek and U. Sheikh and I.G.J. Classen and R.J.E. Jaspers and the TCV team},
	title = {Resonant interaction between runaway electrons and the toroidal magnetic field ripple in {TCV}},
	journal = {Nuclear Fusion},
}
